\documentclass[doublecol,figures]{epl2}
\usepackage{amsmath}
\usepackage{epsfig}
\usepackage{verbatim}
\usepackage{amssymb}
\usepackage{ulem}

\title{Defects and multistability in eutectic solidification patterns}

\author{Andrea Parisi\inst{1,2} \and Mathis Plapp\inst{1}}

\institute{
  \inst{1} Physique de la Mati\`ere Condens\'ee, \'Ecole Polytechnique, CNRS, 91128 Palaiseau, France\\
  \inst{2} Centro de F\'\i{}sica da Mat\'eria Condensada and Departamento de F\'\i{}sica, Faculdade de Ci\^encias, Universidade de Lisboa, Av. Prof. Gama Pinto 2, 1649-003 Lisboa Codex, Portugal
}

\abstract{
We use three-dimensional phase-field simulations to investigate
the dynamics of the two-phase composite patterns formed upon during
solidification of eutectic alloys. Besides the spatially periodic
lamellar and rod patterns that have been widely studied, we find
that there is a large number of additional steady-state patterns
which exhibit stable defects. The defect density can be so high 
that the pattern is completely disordered, and that the distinction 
between lamellar and rod patterns is blurred. As a consequence, 
the transition from lamellae to rods is not sharp, but extends 
over a finite range of compositions and exhibits strong hysteresis.
Our findings are in good agreement with experiments.
}

\begin{document}

\maketitle 

The spontaneous emergence of solidification microstructures during
the freezing of pure substances and alloys is a classic example
for pattern formation outside of equilibrium. These structures
are of considerable practical importance in metallurgy because 
of their influence on the properties of the finished 
material \cite{DantzigRappaz}. Solidification experiments also
provide a particularly well-controlled setting to study fundamental
questions of nonlinear pattern dynamics \cite{Cross93}. 

Eutectic alloys exhibit a triple point in their phase diagram at a 
temperature $T_E$, where liquid of composition $C_E$ can coexist
with two solid phases $\alpha$ and $\beta$ of different compositions
$C_\alpha$ and $C_\beta$. 
Growth from a liquid of composition sufficiently close to $C_E$ 
leads to the formation of a composite solid, with two basic growth
morphologies: alternating parallel platelets (lamellae) of the 
two solid phases, or fibres (rods) of one phase inside a matrix 
of the other, with the centres of the fibres located on a 
triangular lattice. In cross-sections of the finished material, the 
two morphologies look like stripes and round dots, respectively. 
This is reminiscent of the stripe and hexagon patterns that appear in 
many two-dimensional or quasi-two-dimensional systems, both out of 
equilibrium (e.~g. Turing patterns, convection, reaction-diffusion
systems) and in equilibrium (e.~g. thin magnetic films, block 
copolymers). This is of course
not accidental: like in these systems, two antagonistic effects
(here, solute redistribution by diffusion and surface tension)
lead to the emergence of a characteristic length scale.

While it has been known for a long time that rods are preferred over
lamellae when the volume fraction of one of the two phases is small,
the precise conditions for the appearance of these two morphologies,
as well as the nature of the transition between them, have remained
unclear. Here, we show by quantitative three-dimensional phase-field 
simulations of eutectic solidification that the transition 
from rods to lamellae has some features that are not observed 
in other systems. We observe that 
regular arrays of rods exhibit, when their spacing is
increased, a bifurcation from circular rods to symmetry-broken 
rods that have an oval or even dumbbell-like shape. The resulting
stable arrays look like broken lamellae. There is thus a whole
family of spatially periodic states ``in between'' rods and 
lamellae. Furthermore, we find that besides these periodic 
states, a large number of disordered steady-state
configurations exist which consist of a mixture of rods and 
lamellae. Whereas such disordered states have also been found
in other systems, they seem to be particularly stable in
eutectic solidification. As a consequence of the existence
of these numerous intermediate stable states, the transition 
from rods to lamellae is strongly hysteretic. All of these 
observations compare favourably to recent experimental findings.

Let us start out by a brief review of what is known on the
two types of periodic patterns in eutectics. We consider directional 
solidification: a sample is pulled with fixed velocity $V$ 
in an externally imposed temperature gradient $G$. The
liquid freezes into two distinct solid phases of different
compositions. If all crystallographic effects are neglected
and the isotherms are assumed to be planar, the problem is
invariant by rotations and translations in the plane of the
isotherms. Under these conditions, Jackson and Hunt \cite{JH-66} 
obtained two families of approximate steady-state solutions 
for perfectly periodic lamellae and rods, and related the average
front temperature to the lamella or rod spacing $\lambda$.
This front temperature is determined by the interplay between
the diffusive redistribution of alloy components through
the liquid and the effect of surface tension, which shifts the
interface temperature by an amount that is proportional to the 
local interface curvature. For the slow growth regime
that is relevant for most experiments (small P\'eclet number, 
$\lambda V/D\ll 1$, where $D$ is the solute diffusion coefficient 
in the liquid), the global front undercooling with
respect to the eutectic temperature is minimal for a characteristic 
spacing $\lambda_{JH}=KV^{-1/2}$, where the constant $K$ 
depends on the thermophysical properties of the alloy, the volume 
fractions of the phases, and the type of pattern (lamellae or rods).

In order to be observable in experiments, the steady-state solutions
must be stable. Some possible instabilities were already discussed
by Jackson and Hunt \cite{JH-66}. Subsequent research has shown both 
through experiments and  simulations that stable lamellar growth occurs 
within a range of lamellar spacings around $\lambda_{JH}$ both in 
thin \cite{Karma96,Ginibre97,Akamatsu02,Akamatsu04} as well as in extended 
samples \cite{Akamatsu04PRL,Parisi08}; for a more detailed review, 
see \cite{Karma04}. Much less is known about the stability of rods. 
Recent advances have been obtained by using samples of transparent 
organic alloys \cite{Serefoglu08,Perrut09}. Numerically, a few 
examples of rod structures have been calculated by 
phase-field models \cite{Apel02,Lewis04}, but
to our knowledge no detailed quantitative investigation of
the steady-state solutions was previously carried out. We
therefore start our investigations by a detailed survey of
rod stability.

We use an efficient phase-field model recently introduced to simulate 
coupled eutectic growth \cite{FP-03,FP-05}, and used previously to 
investigate the nature of pattern instabilities of lamellar arrays in 
massive samples \cite{Parisi08} as well as the influence of temperature 
profiles on the solidified patterns \cite{Perrut09b}. The model 
has proved to be capable of producing results in excellent agreement 
with experimental observations \cite{FP-05,Parisi08,Perrut09b}.
The equations of the model as well as the method to choose 
appropriate computational parameters are discussed in detail in 
Refs.~\cite{FP-05,Parisi08}. The essential point is that in this
model, the thickness $W$ of the diffuse interfaces can be chosen
freely without introducing computational artefacts as long as it
remains about an order of magnitude smaller that the size of the
smallest features in the pattern.
Since we are interested here in general aspects of eutectic
solidification, we use a eutectic alloy with a symmetric phase
diagram, which yields the best computational 
performance. In this case, all relevant physical parameters
can be grouped to yield three characteristic length scales:
the thermal length $l_T = |m|\Delta C/G$, the diffusion length $l_D=D/V$,
and the capillary length $\bar d= \gamma T_E/(L|m|\Delta C)$, 
with $m$ the liquidus slope, $\Delta C$ the difference in composition 
between the two solids, 
$\gamma$ the solid-liquid surface tension, $T_E$ the eutectic 
temperature, and $L$ the latent heat of melting. For all our 
simulations, we use $l_D/d_0=10^3$ and $l_T/l_D=4$.  

The simulation setup is similar to the one used in Ref.~\cite{Parisi08}.
A rectangular simulation cell is filled by the solid composite up to 
the solid-liquid interface, and by the liquid phase beyond it. 
The isotherms (and hence also the solid-liquid interface) advance 
at constant speed $V$, and the simulation cell is moved whenever the 
front has advanced by a certain amount, so that the interface
is approximately stationary in the cell. An efficient multi-grid 
scheme permits to calculate the concentration field $C(\vec x, t)$ 
up to a very large distance from the interface. An important
parameter is the initial composition $C_0$ of the liquid, because
it determines the equilibrium volume fraction $\eta$ of the $\beta$
phase through the lever rule, $\eta=(C_0-C_\alpha)/\Delta C$. 

\begin{figure}
\onefigure[width=5.5cm]{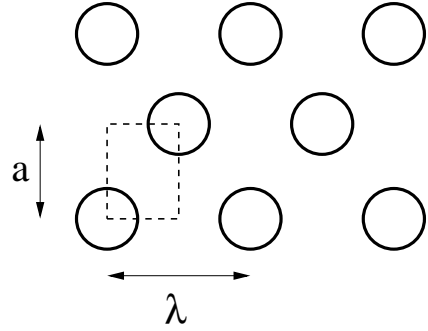}
\caption{\label{fig:simbox}Sketch of the simulation geometry. A perfectly
periodic rod array of horizontal spacing $\lambda$ and distance $a$ between
rows is simulated in the reduced simulation cell indicated by the dashed
rectangle, with reflection boundary conditions.}
\end{figure}

Since it turns out that distortions of the rod array play
an important role for the interpretation of our results, 
it is useful to give some details about the geometry of our
simulation cell, shown in Fig.~\ref{fig:simbox}. We consider
perfectly periodic arrays of rods, where the distance between
rods in a ``horizontal'' row (along the $x$ direction) is 
$\lambda$, and the distance 
between rows is $a$. Steady states of such periodic rod 
arrays can be efficiently simulated by taking advantage of
the two orthogonal planes of mirror symmetry that run 
through the center of each rod. We use the reduced
simulation box indicated by a dashed rectangle, with 
reflection boundary conditions on all lateral boundaries
for all fields. The simulation cell thus contains only two
quarters of rods in two corners. The picture of a full array can 
be reconstructed by adding suitably reflected and shifted copies 
of the simulation cell. The simulation is initialised with two 
circular quarter-rods, where the radius is chosen such that
the volume fractions of the two solid phases are consistent 
with the global alloy composition. The evolution of this
initial state is then followed until a steady state is reached.

A perfect triangular lattice would correspond to $a/\lambda=\sqrt{3}/2$.
However, since our finite-difference discretization uses a regular 
cubic grid, our simulation box has a rational aspect ratio, 
which results in a weakly distorted array. In such an array,
the spacings between rods in a horizontal row and between rods 
in two distinct rows are slightly different. We will use the 
spacing $\lambda$ within a horizontal row in our subsequent plots.
To investigate various spacings at fixed aspect ratio, we 
keep the size of the simulation box in terms of grid points as well 
as the ratio of the grid spacing $\Delta x$ and interface thickness 
$W$ constant (we use $\Delta x/W=0.8$), whereas we vary the ratio 
$W/\bar d$, which corresponds to changing the physical size 
of the simulation box. Typically, we use 32 grid points along
the smaller dimension of the box (of length equal to $\lambda/2$,
see Fig.~\ref{fig:simbox}), which corresponds to a resolution of
$\lambda/W=51.2$, sufficient to guarantee well-converged
results \cite{FP-05,Parisi08}.

\begin{figure}
\onefigure[width=6cm]{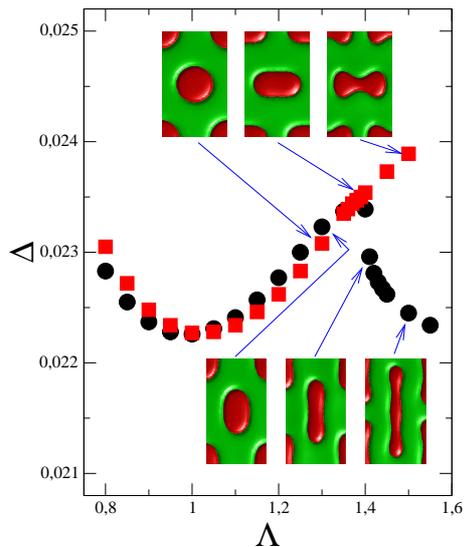}
\caption{\label{fig:jhcurve}Front undercooling $\Delta$ versus 
reduced spacing $\Lambda=\lambda/\lambda_{JH}$ for rod arrays
of two different geometries and volume fraction of the $\beta$ 
phase of $\eta=0.3$. Circles: $a/\lambda=14/16$; squares:
$a/\lambda=13/16$. Insets: reconstructed steady states.}
\end{figure}

\begin{figure}
\onefigure[width=7.0cm]{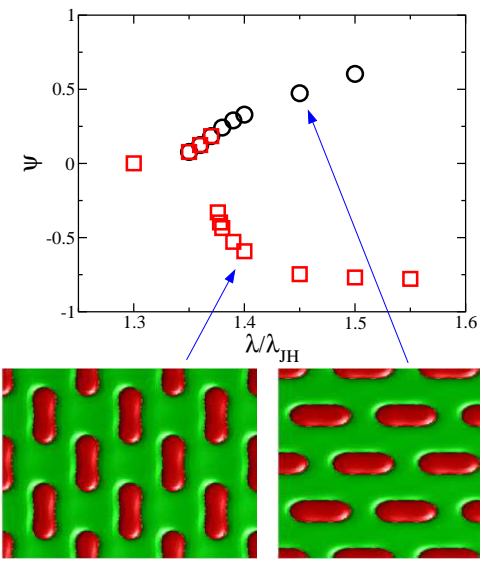}
\caption{\label{fig:bifurc} Bifurcation diagram of the shape transition
for $a/\lambda=13/16$ and $\eta=0.3$.
The parameter $\psi$ is plotted versus the initial rod spacing. Squares
and circles denote runs with different initial conditions: circles,
cylindrical rods; squares, rods elongated towards second neighbours.
The top and bottom branches correspond to rods with a cross-section
elongated towards first and second neighbours, respectively, as shown 
in the pictures of reconstructed arrays (both for $\Lambda=1.4$). 
The centres of the elongated rods remain on the initial rod lattice.}
\end{figure}

We consider first the case of a sample with volume fraction $\eta=0.3$
of the $\beta$ phase (for our symmetric phase diagram, the eutectic
point is at $\eta=0.5$). In Fig.~\ref{fig:jhcurve}, 
we plot the dimensionless undercooling
\begin{equation*}
\Delta = \displaystyle\frac{T_E-T_{\rm front}}{|m|\Delta C}
\end{equation*}
as a function of
the reduced spacing $\Lambda=\lambda / \lambda_{JH}$, where 
$\lambda_{JH}$ is the minimum undercooling spacing calculated 
from the Jackson-Hunt theory for rods. Data for two different
distorted arrays are shown. For an aspect ratio of $a/\lambda=14/16$,
which is slightly larger than $\sqrt{3}/2$,
the undercooling curve exhibits a minimum at a spacing very close to
the one predicted by the Jackson-Hunt calculation, and follows the
expected behaviour $\Delta=\Delta_{\rm min}(\Lambda+1/\Lambda)$ \cite{JH-66}
up to spacings of about $\Lambda=1.3$. However,
as the spacing increases, we observe a change in shape from circular 
rods to elongated rods with the long axis oriented towards the second 
nearest neighbours (see Fig.~\ref{fig:bifurc} for the view of a
reconstructed array). At the same time, the undercooling
starts to differ markedly from the Jackson-Hunt prediction.
For a slightly lower aspect ratio of $a/\lambda=13/16$,
smaller than $\sqrt{3}/2$, the undercooling curve again exhibits
a minimum for a slightly larger spacing, but upon increasing the
spacing, this time the rods elongate in the direction of the 
{\it first} neighbours.

In fact, both of these elongated states can be reached for 
each aspect ratio by starting from initial conditions which 
favour one direction or the other, such as elliptic rods 
elongated towards the first or the second nearest neighbours 
instead of simple circular rods. Thus, the shape transition
is a bifurcation. For a more quantitative description, we define
a shape parameter 
\begin{equation*}
\psi=\displaystyle\frac{d_1-d_2}{d_1+d_2}
\end{equation*}
with $d_1$ and $d_2$ being the diameter of the rod along axes 
oriented towards the first and second nearest neighbours,
respectively. A perfectly circular rod corresponds to 
$\psi=0$, whilst for a deformation towards first (second) 
nearest-neighbours $\psi$ has a positive (negative) sign.
This parameter is plotted versus the spacing in 
Fig.~\ref{fig:bifurc} for $a/\lambda=13/16$. Note that the 
bifurcation is asymmetric because stretching of the rods 
towards first and second nearest neighbours is not equivalent.
Furthermore, for this aspect ratio, the single branch of
weakly distorted rods that exists below the bifurcation
threshold connects continuously to the upper
branch of solutions (horizontally elongated rods), whereas
the second branch appears by a discontinuous jump of the
shape parameter. The opposite behaviour is observed in the
bifurcation diagram for the aspect ratio $a/\lambda=14/16$
(not shown): here, the single branch exhibits slightly
negative values of $\psi$ when it approaches the bifurcation
threshold, and connects continuously to the
lower solution branch (vertically elongated rods).
Thus, both elongated rod states are stable steady-state
solutions, but their respective basins 
of attraction depend on the distortion of 
the initial rod array. Elongation is always favoured in the
direction that is stretched with respect to a symmetric 
hexagonal array.

\begin{figure}
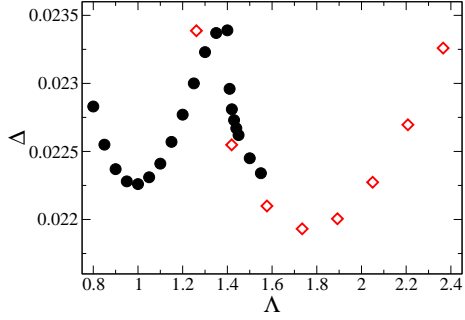

\onefigure[width=6cm]{fig4.eps}
\caption{\label{fig:rodlam}Comparison of the front undercooling 
$\Delta$ of rods with $a/\lambda=14/16$ (circles), and lamellae of half
the original rod spacing (diamonds). In order to achieve a direct 
comparison of corresponding states, for the lamellae the reduced spacing
was calculated as $\Lambda=2\lambda/\lambda_{JH}$, where 
$\lambda_{JH}$ is the {\it rod} minimum undercooling spacing.}
\end{figure}

As can be appreciated from Fig.~\ref{fig:bifurc}, the periodic
arrays of elongated rods become similar, for both orientations, 
to a set of ``broken'' parallel lamellae, with a spacing 
$\lambda_l=\lambda/2$ for the case of $\psi < 0$ and
$\lambda_l=a$ for $\psi > 0$. Indeed, 
this is not just a geometrical similarity: in 
Fig.~\ref{fig:rodlam}, it can be seen that the undercooling 
of a broken lamellar state with $\psi <0$ is very close
to the one for perfect lamellae of half the original spacing.
Note that we have plotted the curve for lamellae as a function
of $2\lambda$ in order to obtain a meaningful comparison in
this graph. The shape change from circular rods to elongated 
rods takes place approximately at the intersection of the 
two Jackson-Hunt curves. This observation provides a simple 
criterion to predict the occurrence of the morphology 
transition, although it gives no direct explanation for 
the mechanism that is at work. Note that we have not observed
any reconnection of the ``broken lamellae'' up to the largest 
spacings investigated ($\Lambda=1.55$). 

The behaviour of the horizontal broken lamellae in the case 
$a/\lambda=13/16$ is quite different. The undercoolings for
lamellae of spacing $a$ are not very different from the
original Jackson-Hunt curve for rods of spacing $\lambda$; the 
undercoolings of the horizontal broken lamellae thus do not exhibit a
dramatic change with the spacing. However, this branch of symmetry-broken 
solution terminates for $\Lambda$ slightly above $1.5$. The 
dumbbell-shaped rods shown in Fig.~\ref{fig:jhcurve} split in 
two and reconnect to form vertical broken lamellae.

\begin{figure}
\onefigure[width=5.6cm]{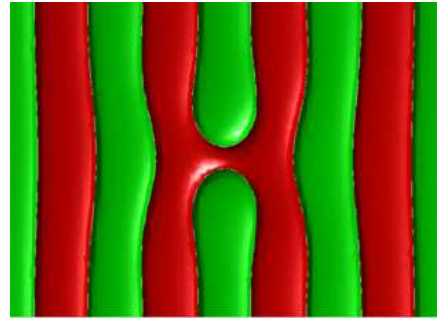}
\caption{\label{fig:defect}Lamellar array with a single defect 
that remains stable. The snapshot shown is a steady state.}
\end{figure}

\begin{figure}
\onefigure[width=5.6cm]{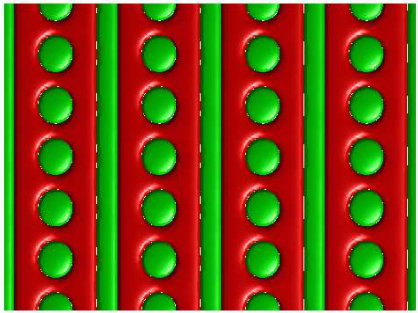}
\caption{\label{fig:cohexistence}Coexistence of rods and 
lamellae. The snapshot shown is a steady state.}
\end{figure}

In summary, the ``broken lamellae'' or ``elongated rod'' 
states have properties that are very close to the ones of
standard lamellae. Therefore, the morphological change
can be interpreted as a transition to a lamellar array
that is not completed, since neighbouring rods do not 
connect. Clearly, a barrier prevents the establishment 
of a simple lamellar pattern.
Further simulations reveal that  the existence of such 
barriers is a general property: whenever defects exist in a 
eutectic solidification pattern, they are very difficult 
to heal. A typical example is shown in Fig.~\ref{fig:defect}.
The initial condition for this simulation is an array of 
parallel lamellae, where a ``cut'' has been made in the
central lamella, thus creating two lamella terminations 
facing each other. The pattern could come back to a
perfect lamellar array by just reconnecting the two
terminations; instead, after an initial transient where 
the two endings slightly adapt in shape and length, the 
simulation evolves towards the steady state shown in 
Fig.~\ref{fig:defect}.
It is thus clear that there exists a large number of steady
states that exhibit stable defects. Indeed, by repeating
the operation of ``cutting'' a lamella, an arbitrary number
of defects can be introduced in a lamellar pattern. Inversely,
if in a perfect rod pattern two rods are ``connected'' to
form a piece of lamella, the latter persists as a stable
steady state. It is even possible to create stable patterns 
which are a complete mix of lamellae and rods, such as shown 
in Fig.~\ref{fig:cohexistence}.

\begin{figure}
\centerline{
\includegraphics[width=2.5cm]{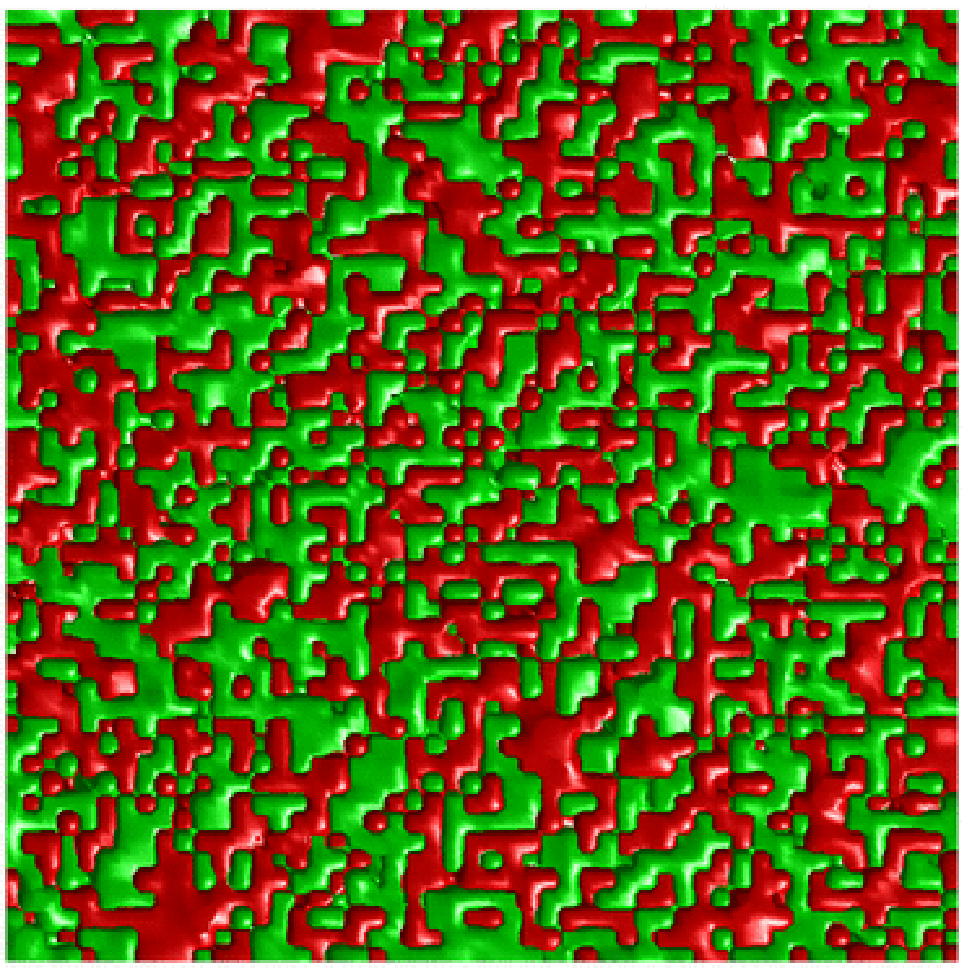}
\includegraphics[width=2.5cm]{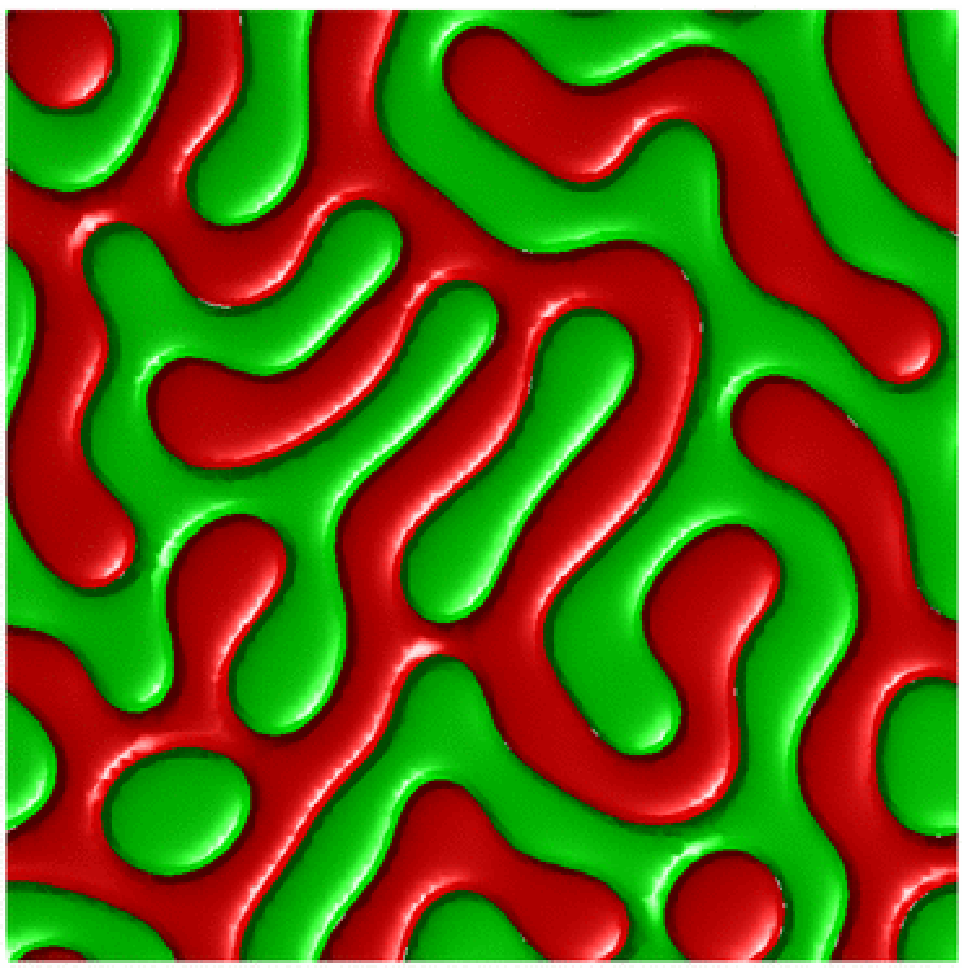}
\includegraphics[width=2.5cm]{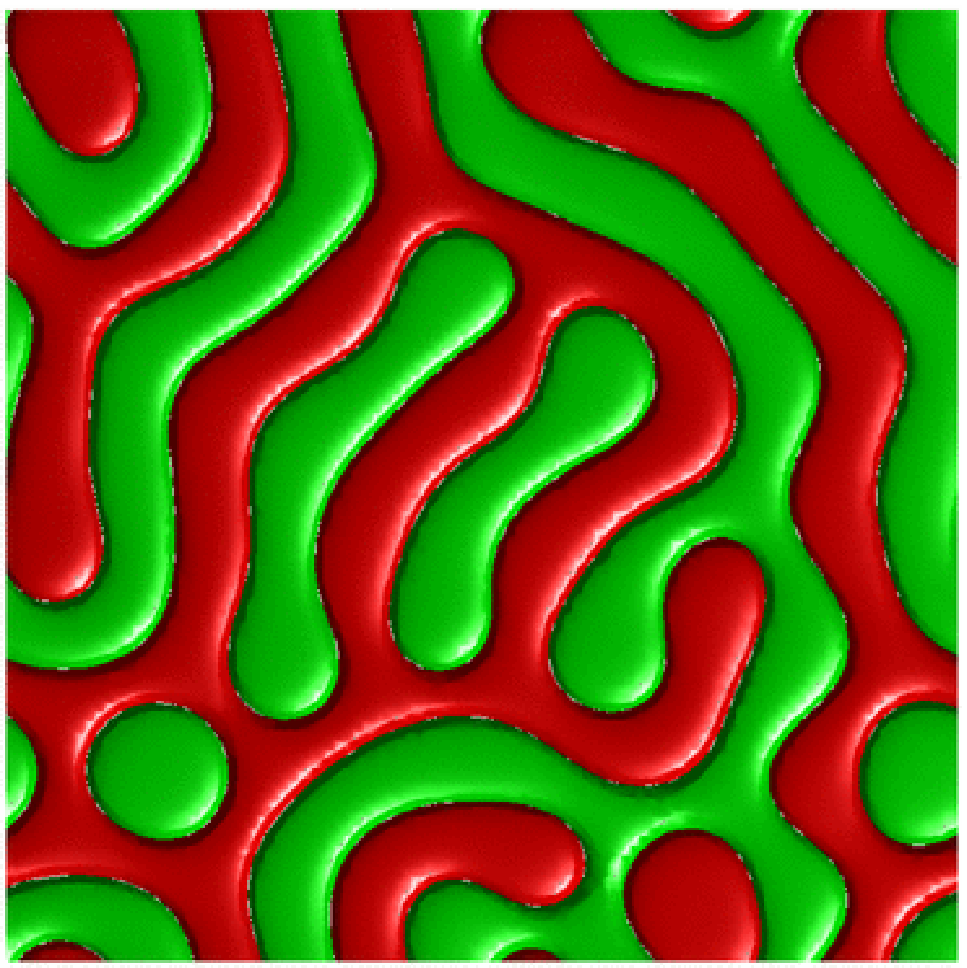}}
\centerline{
\includegraphics[width=2.5cm]{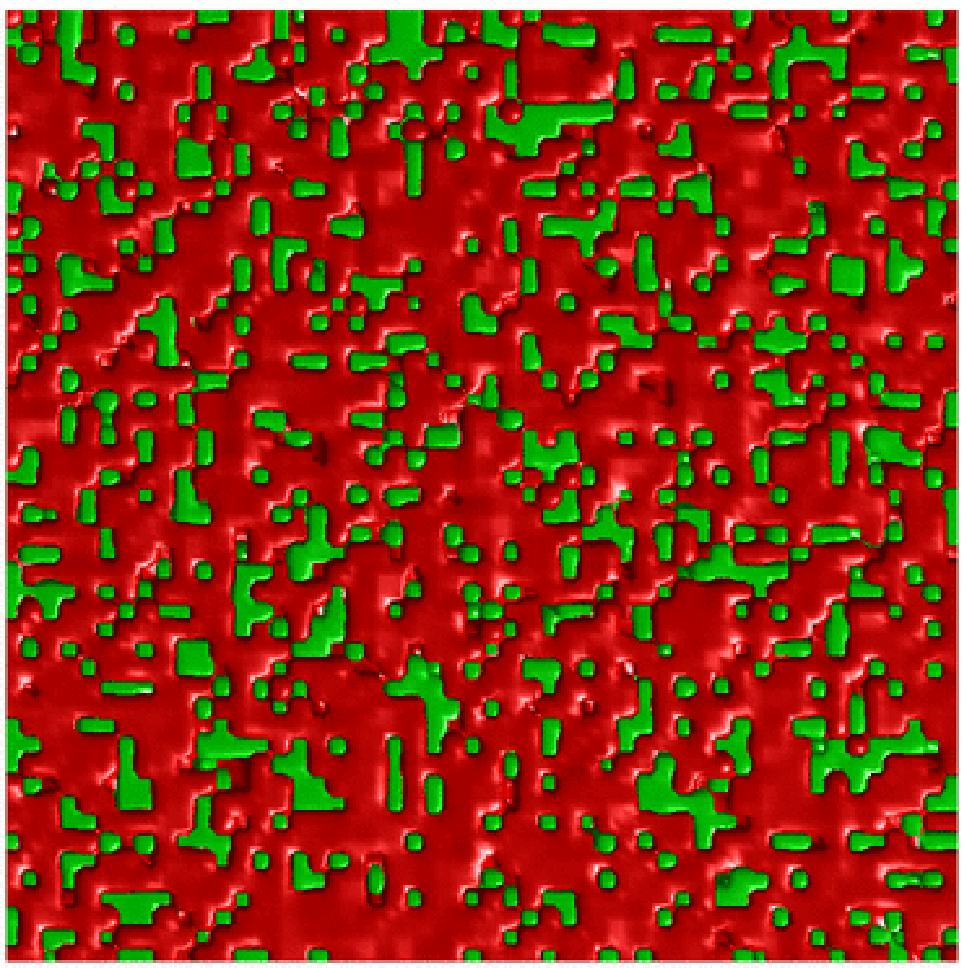}
\includegraphics[width=2.5cm]{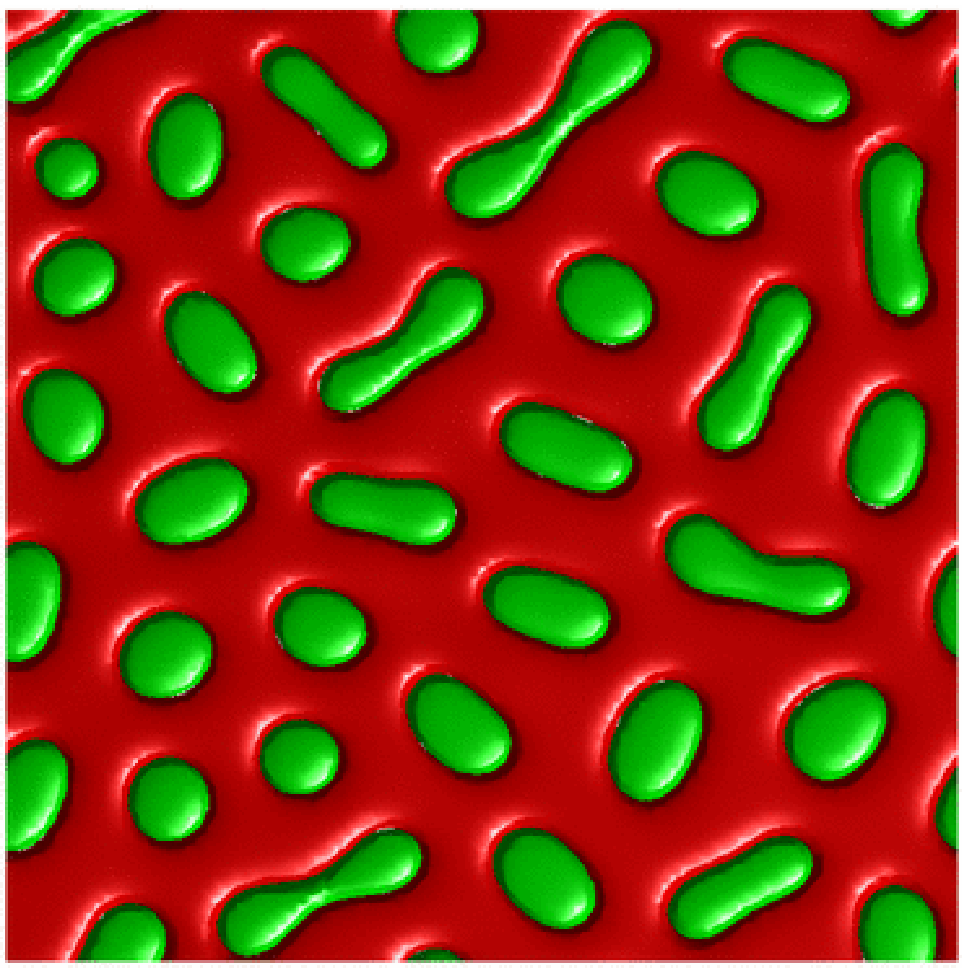}
\includegraphics[width=2.5cm]{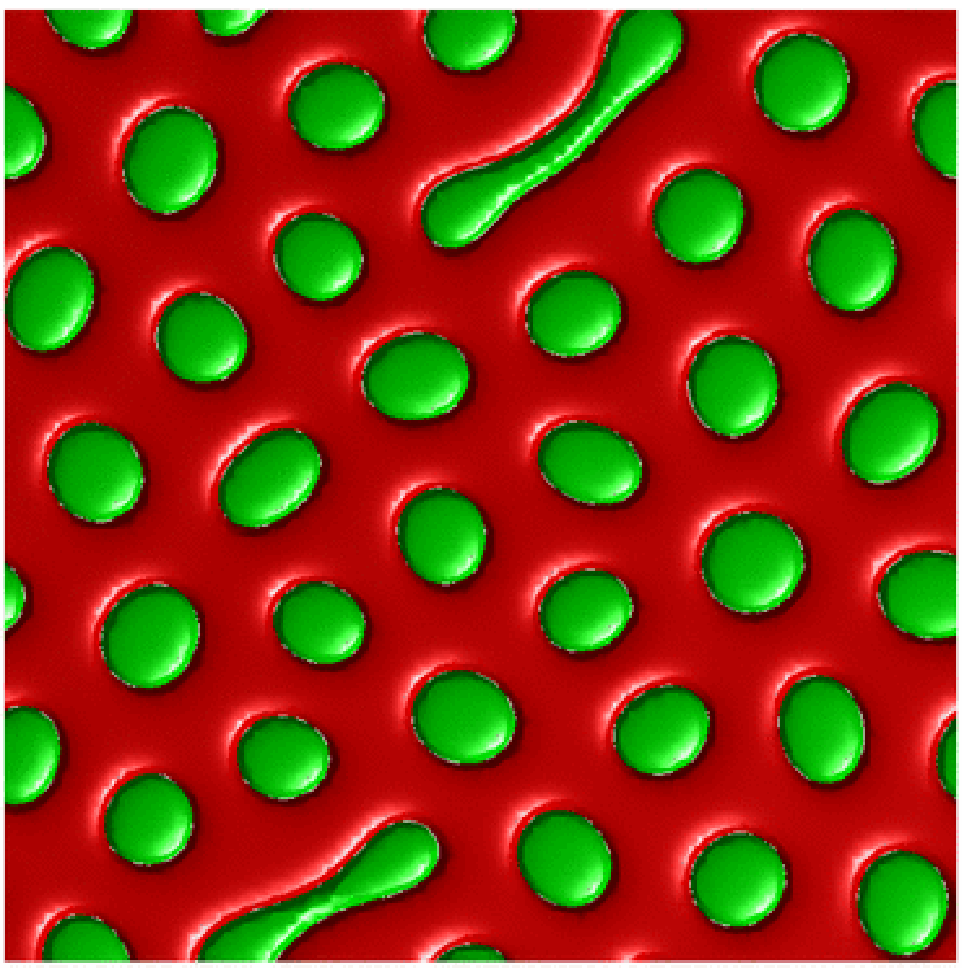}}
\caption{\label{fig:randomfields}(Colour online) Simulations of the 
symmetric model alloy at the eutectic composition ($\eta=0.5$, top
row) and at $\eta=0.3$ (bottom row) in an extended system. The 
initial state is a random distribution of patches of the two 
solid phases. We have used $\bar d/W=0.06218$, and the lateral 
system size is $160\;W$, which corresponds to $1.6\,l_D$ 
and $6.5\,\lambda_{JH}$ (at the eutectic composition). Snapshots 
for both simulations were taken at times $t/t_D=0$, $0.90$ and $21.87$, 
where $t_D=l_D^2/D$ is the diffusion time; the last frame reflects 
a total solidified length of $127.39\,\lambda_{JH}$.}
\end{figure}

The presence of multiple steady states separated by ``barriers'' 
implies that a system started from random initial conditions
will in general not reach a spatially periodic steady state.
Fig.~\ref{fig:randomfields} shows examples of pattern evolution
in an extended system using as initial condition a solid consisting 
of a random distribution of the two phases. For the eutectic
composition, the emerging pattern is a complex network of 
lamellae, the evolution of which is very fast at first:
reconnections between lamellae occur which lead towards a more 
ordered system. However, later on the evolution becomes 
very slow, since it is increasingly difficult for the system 
to overcome the barriers which hinder reconnection. Interestingly, 
the resulting pattern shows, in addition to the emerging 
lamellar pattern, also a few isolated rods, which remain stable
until the end of the simulation. For a $\beta$ volume fraction 
of $\eta=0.3$, mostly rods are formed that are disordered and
elongated at first and then rearrange into a more ordered
pattern. Note, however, that a few isolated ``pieces of lamellae''
remain. A series of similar simulations for various
volume fractions shows that ``pure'' rod patterns are formed
only for volume fractions below $\eta\approx 0.25$.
It should be noted in passing that the final state for $\eta=0.3$ 
mostly consists of rods, although it can be seen from Fig.~\ref{fig:rodlam} 
that the global minimum undercooling is lower for lamellae at 
this composition. It is therefore clear that the criterion
of lowest undercooling, often used in the metallurgical literature,
is not sufficient to predict the emerging morphology.

\begin{figure}
\onefigure[width=7.5cm]{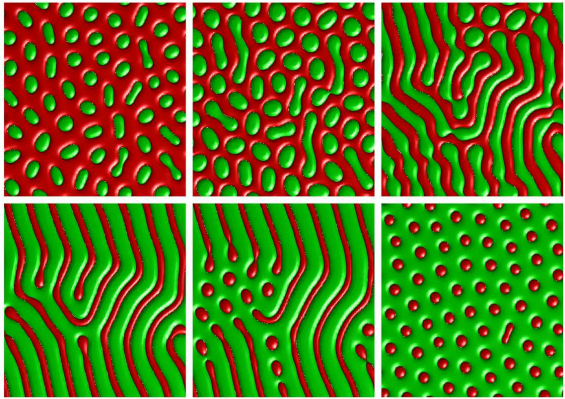}
\caption{\label{fig:bigfigure}(Colour online) Simulation of the symmetric 
model alloy in an extended system, where the volume fraction of $\beta$
phase is varied from $\eta=0.2$ to $\eta=0.8$ over a time of $19.2\,t_D$. 
Snapshots are taken at $t/t_D=3.2$, $8.3$, $10.2$,
$15.4$, $16.7$, and $19.2$; the lateral system size is $1.6\,l_D$, the
same as in Fig.~\ref{fig:randomfields}.}
\end{figure}

These findings imply that the switching between lamellar and 
rod patterns which is observed in real samples can only 
be explained by the presence of a sufficiently large driving 
force. In fig.~\ref{fig:bigfigure} we explore one possibility 
for such global forcing: a change of composition
with time. The simulation is started from a state of highly 
unstable parallel lamellae, which rapidly break up and form 
the irregular pattern of rods shown in the first snapshot.  
The average concentration in the liquid far ahead of the
growth front is changed linearly with time, which implies a change 
in volume fraction from $\eta=0.2$ to $\eta=0.8$ in the course
of the simulation. The figure shows that a transition between 
rod and lamellar patterns can be triggered, but also shows that 
there is a strong hysteresis: the transition from rods to
lamellae occurs at a volume fraction of $\eta\approx 0.5$, whilst the one 
from lamellae to rods occurs at $\eta\approx 0.7$. If the composition was
decreased again starting from the last picture, the transition
from rods to lamellae would happen again at $\eta\approx 0.5$, which 
shows that the patterns found for volume fractions between these 
two limits depend on the history of the system.

It is interesting to notice that both the lamellar and rod
patterns that emerge after the transitions are clearly more
regular than the ones obtained from random initial conditions.
This is a result of a non-trivial spatial organisation of
these transitions. The formation of lamellae occurs by 
successive connections of rods, which could lead to a
random lamellar pattern. However, as was discussed above,
if the initial rod array is not perfectly hexagonal, but
slightly distorted, the shape bifurcation favours a well-defined
direction of elongation, which leads to the formation of
large patches of parallel lamellae. The transition from
lamellae to rods, in turn, starts at lamella terminations 
and propagates along lamellae by successive pinch-offs of
new rods from the ending of the lamella; this leads then
to the breakup of neighbouring lamellae. It would be highly
interesting to study the dynamics of these transitions in
more detail, but this is outside the scope of the present
work.

All our results are in excellent agreement with experimental
observations. Rods that are elongated (both towards first and
second neighbours) have been observed in Al-Cu \cite{Lee05},
mixed states of lamellae and rods have been found both in
Al-Cu \cite{Liu04} and in transparent organic 
alloys \cite{Akamatsu04PRL}. In Fig.~7c of Ref.~\cite{Racek74}, 
alternating lamellae and rows of rods can be seen, which
bear a striking similarity to our Fig.~\ref{fig:cohexistence}.
Furthermore, it was shown that labyrinth states evolve indeed
very slowly in time in the absence of external 
forcings \cite{Perrut09b}. Patterns of coexisting rods
and lamellae were also observed in the peritectic Cu-Sn
alloy \cite{Kohler09}.

Both aspects described above -- shape instabilities of rods
and disordered and ``mixed'' states -- have been observed in
other pattern-forming systems, but there are important differences
between our findings and the ones reported in the literature.
For instance, instabilities leading to dumbbell-shaped domains
have been observed in ferrofluid droplets \cite{Jackson2001}, and 
epitaxial islands \cite{Ni2004}, and have been analysed 
theoretically in systems with a Coulomb interaction \cite{Muratov1997}.
However, in all these systems a long-range repulsion leads to
the fact that the ``tips'' of the domains tend to avoid each other,
instead of forming stripe-like assemblies like in our case.
Furthermore, it is also well know that the transition from rods
to lamellae is hysteretic; in thin magnetic films, this can be
explicitly shown by a calculation of the energies associated
with each pattern \cite{Garel82}. However, in most cases once 
the rods get unstable, they immediately connect to form lamellae, 
the only experimental evidence we are aware of for a (weakly) 
symmetry-broken spot pattern being
Refs.~ \cite{Ouyang1992, Ouyang1991}. Furthermore, coexistence 
between rods and lamellae was observed \cite{Gunaratne1994,Hilali1995}
but coexisting domains of stripes and rods are usually quite
homogeneous, very distinct from the totally disordered and
mixed patterns that we get.

Whereas it is difficult to pinpoint the exact origin of this
difference between eutectic solidification and other pattern-forming
systems, a few ideas can be advanced to motivate further investigation.
First, whereas most of the systems mentioned above can be treated
as quasi-two-dimensional, the eutectic growth front is 
three-dimensional. As mentioned above, one of the main
effects that determine the front structure is capillarity.
Clearly, defects in the pattern are surrounded by a specific local
distribution of curvature, which can make the effective interaction
between them more complicated than the standard attraction or
repulsion generally expected for topological defects. Second, most
of the patterns mentioned above arise when the system undergoes
a bifurcation from a homogeneous to a patterned state 
(e.g. convection). In eutectic solidification, a homogeneous
base state does not exist. Therefore, if the standard phenomenology
of pattern formation applies at all, eutectic solidification 
most likely corresponds to a regime far beyond the bifurcation,
where nonlinear effects are strong. Both of these points can
yield potential explanations for the existence of barriers.

\acknowledgments

We thank S. Akamatsu, S. Bottin-Rousseau, and G. Faivre for 
many discussions. This work was supported by the Centre
National d'Etudes Spatiales (France).

\end{document}